\documentclass{iopconfser}
\usepackage{graphicx}

\newcommand{\eqref}{\eref}
\newcommand{\dd}{\mathrm{d}}
\newcommand{\ee}{\mathrm{e}}

\newcommand{\E}{\mathcal E}

\newcommand{\A}{\mathcal A}
\renewcommand{\H}{{\mathcal H}}

\newcommand{\Y}{{\mathbf Y}}

\renewcommand{\Re}{\mathrm{Re}}

\begin{document}

\title{The balance problem for $n$ aligned black holes}

\author{J\"org Hennig}

\affil{Department of Mathematics and Statistics, University of Otago, Dunedin, New Zealand}

\email{joerg.hennig@otago.ac.nz}
%%%%%%%%%%%%%%%%%%%%%%%%%%%%%%%%%%%%%%%%%%%%%%%%%%%%%%%%%%%%%%%%%%%%%%%%%%%%%
\begin{abstract}
An intriguing open problem in general relativity is whether a stationary equilibrium configuration of multiple, physically relevant black holes can exist. In such a hypothetical setup, the gravitational attraction would need to be balanced by the repulsive spin-spin and electromagnetic interactions. This contribution reports on a method to address this problem for an arbitrary number of $n$ aligned, rotating and possibly charged black holes in an asymptotically flat spacetime. By employing soliton methods to study the underlying boundary value problem for the Einstein-Maxwell equations, we derive the most general form of the boundary data on the symmetry axis. The resulting axis potentials are necessarily rational functions of a specific form, depending on a finite number of parameters. This powerful result reduces the search for $n$-black-hole solutions from solving a highly nonlinear PDE system to analysing a well-defined, finite-parameter family of candidate solutions. We briefly review known results for special cases, such as the constructive uniqueness proofs for a single black hole in vacuum or electrovacuum, and the non-existence proof for two stationary black holes in vacuum, before stating the open problem for more general configurations.
\end{abstract}
%%%%%%%%%%%%%%%%%%%%%%%%%%%%%%%%%%%%%%%%%%%%%%%%%%%%%%%%%%%%%%%%%%%%%%%%%%%%%
\section{Introduction}
~\newline
In Newtonian gravity, the purely attractive nature of the gravitational force precludes the existence of stationary equilibrium configurations of multiple separated bodies. In general relativity, however, the situation is more subtle. The theory's nonlinearity gives rise to effects such as the repulsive spin-spin interaction between co-rotating objects, which, together with electrostatic repulsion for charged bodies, could potentially counteract gravitational attraction. This opens up the possibility that a stationary balance might be achieved. 

Of special significance is the case of black holes, which leads to the fundamental `black hole balance problem'.
While it is known that \emph{static}, reflectionally symmetric $n$-body configurations do not exist \cite{BeigSchoen2009}, the \emph{stationary} case remains an open question. Its resolution is crucial for the topic of black hole uniqueness, as the celebrated uniqueness theorems for the Kerr and Kerr--Newman solutions presuppose a spacetime containing only a single black hole. The existence of stationary multi-black-hole solutions would imply that the end state of gravitational collapse is not necessarily a single Kerr or Kerr-Newman black hole.

We investigate this problem for $n$ aligned, axisymmetric black holes in an asymptotically flat spacetime (i.e.\ with vanishing cosmological constant, $\Lambda=0$). To ensure physical relevance, we restrict our attention to subextremal black holes. This is motivated by the third law of black hole thermodynamics, which posits that extremal black holes, whose horizons have zero surface gravity, cannot be formed in any finite physical process. This is in contrast to the well-known Majumdar-Papapetrou solution \cite{Majumdar1947, Papapetrou1947}, which describes a static equilibrium of multiple, but necessarily extremal, charged black holes.
The physically interesting and unsolved problem thus concerns subextremal configurations.%
\footnote{Recent studies suggest that extremal black holes might be reached dynamically in finite time under the influence of certain exotic matter fields (e.g.\ charged scalar or Vlasov fields), provided we start from suitably fine-tuned initial data \cite{KehleUnger2022, KehleUnger2024}. However, at this stage, the consensus for more astrophysically relevant models remains that only subextremal black holes are stable physical end-states.}
%%%%%%%%%%%%%%%%%%%%%%%%%%%%%%%%%%%%%%%%%%%%%%%%%%%%%%%%%%%%%%%%%%%%%%%%%%%%%
\section{Mathematical Formulation and Soliton Methods}
~\newline
A stationary and axisymmetric spacetime is endowed with commuting Killing vector fields $\partial_t$ and $\partial_\varphi$. For such spacetimes, we can introduce Weyl--Lewis--Papapetrou coordinates $(\rho, \zeta, \varphi, t)$, in which the line element takes the form
\begin{equation}\label{eq:metric}
 \dd s^2=f^{-1}
          \left[\ee^{2k}(\dd\rho^2+\dd\zeta^2)+\rho^2\,\dd\varphi^2\right]
         -f(\dd t+a\,\dd\varphi)^2.
\end{equation}
Here, the metric functions $f, k, a$ depend only on the coordinates $\rho$ and $\zeta$. For electrovacuum spacetimes, the Einstein--Maxwell equations can be elegantly reformulated as the Ernst equations \cite{Ernst1968b} for two complex potentials, the gravitational Ernst potential $\E(\rho,\zeta)$ and the electromagnetic Ernst potential $\Phi(\rho,\zeta)$,
\begin{equation}\label{eq:Ernst}
 f\Delta\E   = (\nabla\E+2\bar\Phi\nabla\Phi)\cdot\nabla\E,\quad
  f\Delta\Phi = (\nabla\E+2\bar\Phi\nabla\Phi)\cdot\nabla\Phi,
\end{equation}
where $f=\Re(\E)+|\Phi|^2$. 

A key insight is that this nonlinear system is completely integrable, i.e.\ it belongs to the remarkable class of soliton equations. For a more detailed introduction to the application of soliton methods in this context, see \cite{Hennig2025}.
This means it is equivalent to the integrability condition of an associated linear matrix problem (LP) \cite{Belinski1979, NeugebauerKramer1983}. This LP is a system of linear first-order PDEs for a $3\times3$ matrix function $\Y(\rho,\zeta;K)$, which depends on the spacetime coordinates and an auxiliary complex `spectral' parameter~$K$.
The explicit form of the LP can be found in \cite{Meinel2012, Hennig2020, Hennig2025}.

A configuration of $n$ aligned black holes is then described by a boundary value problem for the Ernst equations. In the chosen coordinates, the event horizons $\H_i$ of the black holes --- while being topological spheres --- are represented as $n$ disjoint intervals on the $\zeta$-axis ($\rho=0$). These are separated by segments of the symmetry axis $\A_j$. The task is to solve the LP in the exterior domain $\rho\ge0$ subject to physical boundary conditions: regularity on the axis segments, specific conditions on the horizons reflecting their nature as null surfaces rotating with constant angular velocities $\Omega_i$, and asymptotic flatness at infinity.
%%%%%%%%%%%%%%%%%%%%%%%%%%%%%%%%%%%%%%%%%%%%%%%%%%%%%%%%%%%%%%%%%%%%%%%%%%%%%
\section{Axis Potentials and Main Result}
~\newline
While solving the LP in the full domain appears intractable, sufficient information can be extracted by integrating it along the boundaries. On the $\zeta$-axis, where $\rho=0$, the LP simplifies to a system of ordinary differential equations, which can be solved exactly on the axis parts $\A_j$ and horizons $\H_i$. The solutions involve $K$-dependent integration `constants' in the form of $3\times3$ matrices. These matrices turn out not to be independent, but rather are constrained by continuity conditions imposed at the poles of the black holes (i.e.\ at the endpoints of the horizon intervals) and by the asymptotic conditions at infinity.

The derivation crucially relies on two structural properties of the LP. Firstly, the solution $\Y$, as a function of $K$, is defined on a two-sheeted Riemannian surface due to a square-root dependence in the LP's coefficients. A simple algebraic transformation relates the solutions on the two sheets. This is essential for connecting the solution as $\zeta\to\infty$ and $\zeta\to-\infty$ at spatial infinity. Secondly, the transformation to a frame co-rotating with one of the black holes (which is necessary to correctly implement the horizon boundary conditions) can also be expressed as a simple matrix multiplication acting on $\Y$.

By combining the continuity conditions and exploiting these structural properties, we can determine the functional form of the Ernst potentials on the symmetry axis. The final constraint arises from the requirement that the solution $\Y$ be single-valued at the confluent branch points of the spectral function on the axis. This leads to a set of algebraic constraints on the potentials themselves. A detailed analysis of these constraints yields our main result \cite{Hennig2020}:

\emph{
If a stationary equilibrium configuration of $n$ aligned, rotating, and possibly charged black holes exists, its Ernst potentials on the uppermost part of the symmetry axis, $\A_1$, must necessarily have the rational form}
\begin{equation}\label{eq:Ernstformulae}
 \E(\zeta)   = \frac{\pi_n(\zeta)}{r_n(\zeta)},\quad
 \Phi(\zeta) = \frac{\pi_{n-1}(\zeta)}{r_n(\zeta)}.
\end{equation}
Here, $\pi_n$ and $r_n$ are monic complex polynomials of degree $n$, and $\pi_{n-1}$ is a complex polynomial of degree $n-1$. The Hauser-Ernst theorem \cite{HauserErnst1981} guarantees that these axis data uniquely determine the solution in the entire spacetime. Other soliton methods can then be used to obtain them explicitly. 

This result thus achieves a dramatic simplification: The infinite-dimensional problem of solving a PDE system is reduced to the finite-dimensional problem of finding suitable coefficients of these polynomials for which physically acceptable configurations are obtained.
%%%%%%%%%%%%%%%%%%%%%%%%%%%%%%%%%%%%%%%%%%%%%%%%%%%%%%%%%%%%%%%%%%%%%%%%%%%%%
\section{Discussion and Known Cases}
~\newline
The rational structure (\ref{eq:Ernstformulae}) imposes a necessary condition for the existence of $n$-black-hole configurations. However, an arbitrary choice of polynomial coefficients will generally lead to an unphysical solution plagued by singularities or other defects. A physically viable solution must satisfy several further regularity conditions:
\begin{itemize}
    \item Freedom from naked singularities off the axis.
    \item Absence of a NUT parameter to ensure the correct asymptotic behaviour.
    \item Absence of conical singularities (or `struts') on the axis segments between the black holes.
    \item Vanishing net magnetic charge.
\end{itemize}
Verifying these conditions for the general case is a formidable analytical challenge. However, for some special cases, the problem has been fully resolved.

\paragraph{Case $n=1$ (single black hole):}
For $n=1$ in vacuum ($\Phi=0$), the Ernst potential on the axis must be $\E(\zeta) = (\zeta+c_1)/(\zeta+c_2)$. The regularity conditions uniquely fix the constants in terms of the black hole's mass and angular momentum, leading to the known axis potential for the Kerr solution. A similar procedure in electrovacuum yields the Kerr-Newman solution. This framework thus provides constructive proofs of the uniqueness of these fundamental solutions \cite{NeugebauerMeinel2003, Meinel2012}.

\paragraph{Case $n=2$ (vacuum):}
For two uncharged black holes, the axis potential must be a ratio of two quadratic polynomials. This uniquely identifies the double-Kerr-NUT family of solutions \cite{KramerNeugebauer1980} as the sole candidate. A subsequent, detailed analysis showed that, for any choice of parameters within this family that is free of struts, at least one of the two black holes violates the universal inequality $8\pi|J|<\mathcal{A}$, which is a necessary condition for any regular subextremal black hole with angular momentum $J$ and horizon area $\mathcal{A}$ \cite{HennigAnsorgCederbaum2008}. This constitutes a rigorous proof of the \emph{non-existence} of stationary two-black-hole configurations in vacuum \cite{NeugebauerHennig2009, Chrusciel2011}.

\paragraph{Open Problems:}
The balance problem remains open for two charged black holes in electrovacuum, and for three or more black holes in either vacuum or electrovacuum. For each $n$, our result provides a well-defined, finite-parameter family of candidate solutions. Future work must decide whether parameter choices within these families can satisfy all physical requirements simultaneously, or if, as in the two-black-hole vacuum case, subtle pathologies will always rule out such intriguing configurations.

%%%%%%%%%%%%%%%%%%%%%%%%%%%%%%%%%%%%%%%%%%%%%%%%%%%%%%%%%%%%%%%%%%%%%%%%%%%%%

\end{document}